\pdfoutput=1
\newif\ifplainstyle
\plainstyletrue
\plainstylefalse

\newif\ifjhepstyle
\jhepstyletrue
\newif\ifprstyle
\prstyletrue
\prstylefalse
\ifprstyle
	\documentclass[twocolumn,nofootinbib]{revtex4-1}
\else
	\documentclass[11pt,a4paper]{article}
\fi
\ifjhepstyle
	\usepackage{jheppub}
	\usepackage{amsfonts}
	\usepackage{verbatim}
	\usepackage{float}
	\usepackage{color}
	\usepackage{array}
	\makeatletter
	\def\@fpheader{\phantom{:-)}}
	\makeatother
\else	
 	\ifprstyle
		\usepackage{verbatim}
		\usepackage{amsmath,amsfonts,amssymb}
		\usepackage[colorlinks=true
                	,urlcolor=blue
                	,anchorcolor=blue
                	,citecolor=blue
                	,filecolor=blue
                	,linkcolor=blue
                	,menucolor=blue
                	]{hyperref}
	\else
                \makeatletter
                \normalsize
                \divide\@tempdima\baselineskip
                \@tempcnta=\@tempdima
                \skip\@mpfootins = \skip\footins
                \renewcommand\section{\@startsection{section}{1}{\z@}%
                                                   {-3.5ex \@plus -1.3ex \@minus -.7ex}%
                                                   {2.3ex \@plus.4ex \@minus .4ex}%
                                                   {\normalfont\large\bfseries}}
                \renewcommand\subsection{\@startsection{subsection}{2}{\z@}%
                                                   {-2.3ex\@plus -1ex \@minus -.5ex}%
                                                   {1.2ex \@plus .3ex \@minus .3ex}%
                                                   {\normalfont\normalsize\bfseries}}
                \renewcommand\subsubsection{\@startsection{subsubsection}{3}{\z@}%
                                                   {-2.3ex\@plus -1ex \@minus -.5ex}%
                                                   {1ex \@plus .2ex \@minus .2ex}%
                                                   {\normalfont\normalsize\bfseries}}
                \renewcommand\paragraph{\@startsection{paragraph}{4}{\z@}%
                                                   {1.75ex \@plus1ex \@minus.2ex}%
                                                   {-1em}%
                                                   {\normalfont\normalsize\bfseries}}
                \renewcommand\subparagraph{\@startsection{subparagraph}{5}{\z@}%
                                                   {1.75ex \@plus1ex \@minus .2ex}%
                                                   {-1em}%
                                                   {\normalfont\normalsize\itshape}}
                
                \renewcommand{\@dotsep}{10000}
                
                \def\fnum@figure{\textbf{\figurename\nobreakspace\thefigure}}
                \def\fnum@table{\textbf{\tablename\nobreakspace\thetable}}
                
                \long\def\@makecaption#1#2{%
                  \vskip\abovecaptionskip
                  \sbox\@tempboxa{\small #1. #2}%
                  \ifdim \wd\@tempboxa >\hsize
                    \small #1. #2\par
                  \else
                    \global \@minipagefalse
                    \hb@xt@\hsize{\hfil\box\@tempboxa\hfil}%
                  \fi
                  \vskip\belowcaptionskip}

                \renewenvironment{thebibliography}[1]{%
                \begin{oldthebibliography}{#1}%
                \small%
                \raggedright%
                \setlength{\itemsep}{5pt plus 0.2ex minus 0.05ex}%
                }%
                {%
                \end{oldthebibliography}%
                }
                \makeatother

            	\usepackage{verbatim}
            	\usepackage{cite}
            	\usepackage{setspace}
            	\usepackage{amsmath,amsfonts,amssymb}
            	\usepackage[colorlinks=true
            	,urlcolor=blue
            	,anchorcolor=blue
            	,citecolor=blue
            	,filecolor=blue
            	,linkcolor=blue
            	,menucolor=blue
            	,linktoc=page
            	]{hyperref}
            	\usepackage{float}
            	\restylefloat{table}
            	\renewcommand{\arraystretch}{1.5}
            	\numberwithin{equation}{section}
	\fi
\fi

\usepackage{subfig}

\usepackage{etoolbox}
\makeatletter
\def\hlinewd#1{%
\noalign{\ifnum0=`}\fi\hrule \@height #1 %
\futurelet\reserved@a\@xhline}
\makeatother

\allowdisplaybreaks

\usepackage{multirow}

\usepackage[normalem]{ulem}
\usepackage{mathtools}
\usepackage{bbold}
\usepackage{transparent}
\usepackage{bm}
\usepackage{enumitem}
\usepackage[textsize=scriptsize,textwidth=2.45cm]{todonotes}

\long\def\symfootnote[#1]#2{\begingroup%
\def\thefootnote{\fnsymbol{footnote}}\footnote[#1]{#2}\endgroup}

\def\({\left (}
\def\){\right )}

\def\p{\partial}

\def\l{\lambda}

\newcommand{\eq}[1]{\begin{equation}#1\end{equation}}
\newcommand{\eqst}[1]{\begin{equation*}#1\end{equation*}}
\newcommand{\eqa}[1]{\begin{align}#1\end{align}}

\newcommand{\eqsp}[1]{\begin{equation}\begin{split}#1\end{split}\end{equation}}

\makeatletter
\let\save@mathaccent\mathaccent
\newcommand*\if@single[3]{%
  \setbox0\hbox{${\mathaccent"0362{#1}}^H$}%
  \setbox2\hbox{${\mathaccent"0362{\kern0pt#1}}^H$}%
  \ifdim\ht0=\ht2 #3\else #2\fi
  }
\newcommand*\rel@kern[1]{\kern#1\dimexpr\macc@kerna}
\newcommand*\widebar[1]{\@ifnextchar^{{\wide@bar{#1}{0}}}{\wide@bar{#1}{1}}}
\newcommand*\wide@bar[2]{\if@single{#1}{\wide@bar@{#1}{#2}{1}}{\wide@bar@{#1}{#2}{2}}}
\newcommand*\wide@bar@[3]{%
  \begingroup
  \def\mathaccent##1##2{%
    \let\mathaccent\save@mathaccent
    \if#32 \let\macc@nucleus\first@char \fi
    \setbox\z@\hbox{$\macc@style{\macc@nucleus}_{}$}%
    \setbox\tw@\hbox{$\macc@style{\macc@nucleus}{}_{}$}%
    \dimen@\wd\tw@
    \advance\dimen@-\wd\z@
    \divide\dimen@ 3
    \@tempdima\wd\tw@
    \advance\@tempdima-\scriptspace
    \divide\@tempdima 10
    \advance\dimen@-\@tempdima
    \ifdim\dimen@>\z@ \dimen@0pt\fi
    \rel@kern{0.6}\kern-\dimen@
    \if#31
      \overline{\rel@kern{-0.6}\kern\dimen@\macc@nucleus\rel@kern{0.4}\kern\dimen@}%
      \advance\dimen@0.4\dimexpr\macc@kerna
      \let\final@kern#2%
      \ifdim\dimen@<\z@ \let\final@kern1\fi
      \if\final@kern1 \kern-\dimen@\fi
    \else
      \overline{\rel@kern{-0.6}\kern\dimen@#1}%
    \fi
  }%
  \macc@depth\@ne
  \let\math@bgroup\@empty \let\math@egroup\macc@set@skewchar
  \mathsurround\z@ \frozen@everymath{\mathgroup\macc@group\relax}%
  \macc@set@skewchar\relax
  \let\mathaccentV\macc@nested@a
  \if#31
    \macc@nested@a\relax111{#1}%
  \else
    \def\gobble@till@marker##1\endmarker{}%
    \futurelet\first@char\gobble@till@marker#1\endmarker
    \ifcat\noexpand\first@char A\else
      \def\first@char{}%
    \fi
    \macc@nested@a\relax111{\first@char}%
  \fi
  \endgroup
}
\makeatother
\newcommand{\ThisIsTheTitle}{The on-shell action of supergravity  \& \\ the $\bm{B}$-side of TsT and single-trace $\bm{T\bar T}$}
\newcommand{\ThisIsAuthorOne}{Luis Apolo}
\newcommand{\ThisIsEmailOne}{apolo@bimsa.cn}
\newcommand{\TheseAreTheKeywords}{}
\newcommand{\ThisIsTheAbstract}{
We propose boundary terms for the action of the NS sector of supergravity on $M_3 \times S^3 \times T^4$ spacetimes --- where $M_3$ is an AdS$_3$ or a linear dilaton background --- that render the Brown-York stress tensor and the on-shell action finite. For AdS$_3$ backgrounds, we show that the on-shell action yields a free energy with chemical potentials determined by the $B$-field. TsT (T-duality + shift + T-duality) transformations of these backgrounds generate classes of linear dilaton backgrounds distinguished by their boundary conditions. Among them, there is one class of backgrounds that reproduces features of the single-trace $T\bar T$ deformation. These backgrounds have additional chemical potentials that can be turned off by large gauge transformations of the $B$-field. We show that the Brown-York stress tensor of these backgrounds reproduces the energies and the trace flow equation of $T\bar T$-deformed CFTs. We also show that the on-shell action matches the $T\bar T$ partition function and discuss the interpretation of these results.}
\ifjhepstyle
        \title{\ThisIsTheTitle}
  
        \author[]{\ThisIsAuthorOne}
        \affiliation[]{Beijing Institute of Mathematical Sciences and Applications, Beijing 101408, China}
               
        \emailAdd{\ThisIsEmailOne}
        \abstract{\ThisIsTheAbstract} 
\fi
\newcommand{\CTtitle}[1]{\vbox{\center\bf\LARGE{#1}}}
\newcommand{\CTauthor}[1]{\vbox{\center{#1}}}
\newcommand{\CTaddress}[1]{\vbox{\center\footnotesize\it{#1}}}
\newcommand{\CTemail}[1]{\vbox{\center\footnotesize\texttt{#1}}}
\def\vac{vac}
\def\bh{bh}
\def\bshift{\gamma}
\def\Bshift{\Gamma}
\begin{document}

%%%%%%%%%%%%%%%%%%%%%%%%%%%%%%%%%%%%%%%%%%%%%%
\ifjhepstyle
       \maketitle
        \flushbottom
        %\raggedbottom
\else
        \begin{center} 
        \thispagestyle{empty}
        \phantom{a}
        \vspace{100pt}
        
        \CTtitle{\ThisIsTheTitle}
         
        \vspace{10pt}
       
        \CTauthor{\ThisIsAuthorOne}
        
        \vspace{3pt}
        
        %{\footnotesize\it } 
        \CTaddress{Beijing Institute of Mathematical Sciences and Applications, Beijing 101408, China}
        
        \vspace{6pt}
        
        %\vbox{\footnotesize{ \texttt {\ThisIsEmailOne}}} 
        \CTemail{\ThisIsEmailOne}
        
        \end{center}
        
        \vspace{5.5pt}
        
        \begin{abstract}
        \vspace{-1.5pt}
        \noindent \ThisIsTheAbstract

        \end{abstract}
        
        \vspace{30pt}
       
        \newpage
         
        \noindent\hrule
        \tableofcontents
        
        \bigskip 
	\bigskip 
	\noindent\hrule
	\bigskip
\fi
%%%%%%%%%%%%%%%%%%%%%%%%%%%%%%%%%%%%%%%%%%%%%%

%%%%%%%%%%%%%%%%%%%%%%%%%%%%%%%%%%%%%%%%%%%%%%
\section{Introduction}
%%%%%%%%%%%%%%%%%%%%%%%%%%%%%%%%%%%%%%%%%%%%%%

The $T\bar T$ deformation --- an irrelevant deformation built from the components of the stress tensor --- has received considerable attention due to its solvability \cite{Zamolodchikov:2004ce, Smirnov:2016lqw,Cavaglia:2016oda} and connections to two-dimensional theories of quantum gravity \cite{Dubovsky:2012wk,Dubovsky:2013ira,Dubovsky:2017cnj,Cardy:2018sdv,Callebaut:2019omt,Tolley:2019nmm} (see \cite{Jiang:2019epa,Guica:2020rev,He:2025ppz} for reviews). The solvability of the $T\bar T$ deformation makes it particularly appealing in extending holography beyond the AdS/CFT correspondence. For a generic CFT$_2$, the $T\bar T$ operator modifies the boundary conditions of the metric without changing the locally AdS$_3$ nature of the spacetime \cite{Guica:2019nzm}. In certain scenarios, these boundary conditions can be interpreted as a cutoff in the bulk whose location is controlled by the $T\bar T$ deformation parameter \cite{McGough:2016lol}. 

The $T\bar T$ deformation is also capable of changing the bulk geometry, leading to a  model of holography for non-AdS spacetimes. This model is obtained by deforming string theory on AdS$_3$ by a marginal operator on the worldsheet that corresponds to  the single-trace $T\bar T$ operator in the dual CFT \cite{Giveon:2017nie}.\footnote{The single-trace $T\bar T$ deformation of a symmetric product orbifold CFT consists of deforming each copy of the CFT by the $T\bar T$ operator in that copy.} The deformation results in linear dilaton backgrounds that reproduce several features of single-trace $T\bar T$-deformed CFTs \cite{Giveon:2017nie, Hashimoto:2019wct, Apolo:2019zai, Hashimoto:2019hqo, Georgescu:2022iyx,Apolo:2023aho,Chakraborty:2023wel,Cui:2023jrb,Du:2024bqk}. A more precise statement of the correspondence has been made in the tensionless limit of string theory on $M_3 \times S^3 \times T^4$, where $M_3$ denotes the linear dilaton backgrounds. In this case, the holographic dual is identified with the single-trace $T\bar T$ deformation of the symmetric product orbifold of $T^4$ \cite{Dei:2024sct}. 

The linear dilaton backgrounds mentioned above can be obtained via TsT (T-duality + shift + T-duality) transformations of AdS$_3$ \cite{Araujo:2018rho,Borsato:2018spz,Apolo:2019zai}. The reason is that TsT transformations are equivalent to exactly marginal deformations of the string worldsheet that generalize the one considered in \cite{Giveon:2017nie}. Using TsT transformations, it is possible to construct the space of linear dilaton backgrounds with arbitrary energy and angular momentum. The solution space includes, in particular, the ground state and rotating black holes \cite{Apolo:2019zai}, backgrounds that can also be obtained by other approaches as described in \cite{Chakraborty:2020swe, Chakraborty:2023mzc, Chakraborty:2023zdd} and \cite{Chang:2023kkq}. 

We will see that the linear dilaton backgrounds relevant for single-trace $T\bar T$ holography generated by TsT have nonvanishing chemical potentials. For the black holes, the chemical potential has been identified with the value of the $B$-field at the horizon \cite{Compere:2007vx,Apolo:2019zai}. These chemical potentials are not fixed in supergravity, and do not affect the gravitational charges, entropy, or thermodynamics of these solutions --- quantities that match those of (single-trace) $T\bar T$-deformed CFTs. We will show that this is a consequence of keeping the conjugate electric charge fixed in the space of solutions. The situation is different in string theory, where the $B$-field has been argued to be fixed uniquely in AdS$_3$ and the linear dilaton backgrounds \cite{Ashok:2021ffx,Martinec:2023plo}, leading to a specific choice of chemical potentials. Reproducing these chemical potentials requires shifts of the $B$-field after TsT \cite{Chakraborty:2024ugc,Giveon:2024sgz}, and it would be interesting to understand how these shifts can be determined in supergravity. 

In this paper, we will probe the effect of the chemical potentials and the $B$-field by evaluating the on-shell action of supergravity. The latter is known to vanish, up to boundary terms, due to the equation of motion of the dilaton (see e.g.~\cite{Kurlyand:2022vzv}). Here, we propose a set of boundary terms that render the Brown-York stress tensor and the on-shell action of the AdS$_3$ and linear dilaton backgrounds finite, with a well-defined variational principle that fixes the electric charge and the asymptotic value of the metric. These boundary terms consist of the Gibbons-Hawking term, a counterterm given by the area of the induced metric $\gamma_{\mu\nu}$ at the asymptotic boundary $\p M$, and two terms depending on the field strength $H = dB$ that are given by
\eq{\label{boundaryterms_intro}
I_{ct_2} = - \frac{1}{4\pi \ell_s^4} \int_{\p M} e^{-2\Phi} \iota_n H,   \qquad I_{B} = \frac{1}{4\pi \ell_s^4} \int_{\p M} e^{-2\Phi} \iota_n H \wedge \star_2 B,
}
where $n^\mu$ is the outward-pointing vector normal to $\partial M$. The first of these terms is a counterterm that renders the on-shell action finite. In asymptotically AdS$_3$ spacetimes, this term is proportional to the counterterm mentioned above when evaluated on shell. The second term is equivalent on shell to the one proposed by Kurlyand and Tseytlin \cite{Kurlyand:2022vzv}, up to large gauge transformations of the $B$-field. This term was also considered recently in \cite{Detournay:2024lhz}, where it was shown to fix the electric charge. 

The on-shell action of supergravity reduces to a collection of boundary terms, of which only the terms in \eqref{boundaryterms_intro} are not TsT invariant. We will evaluate this action on the AdS$_3$ and linear dilaton backgrounds. For the AdS$_3$ backgrounds, we find that the Euclidean on-shell action yields the free energies of thermal AdS/BTZ black holes with additional chemical potentials. The latter correspond to the values of the $B$-field at the origin/horizon. The value of the $B$-field has been proposed to be fixed in string theory \cite{Ashok:2021ffx,Martinec:2023plo} (see also \cite{Hemming:2001we,Troost:2002wk,Hemming:2002kd,Rangamani:2007fz,Nippanikar:2021skr}) and to be nonzero at the horizon.  
%Different proposals for the value of the $B$-field in string theory on AdS$_3$ have been put forward in the literature \cite{Hemming:2001we,Troost:2002wk,Hemming:2002kd,Rangamani:2007fz,Ashok:2021ffx,Nippanikar:2021skr,Martinec:2023plo}. 
From the point of view of supergravity, we will see that a vanishing $B$-field at the origin/horizon yields a free energy that reproduces the universal partition function of two-dimensional CFTs in the semiclassical limit \cite{Hartman:2014oaa}.

The TsT transformation of the AdS$_3$ backgrounds is sensitive to the presence of the chemical potentials described above, the latter of which affect the asymptotic behavior of the metric and the $B$-field. We find that the linear dilaton backgrounds related to the single-trace $T\bar T$ deformation can only be obtained for a specific and nonvanishing choice of these potentials. For this choice, which agrees with the one made in the study of TsT transformations and other irrelevant deformations \cite{Apolo:2018qpq,Apolo:2019yfj,Apolo:2019zai,Apolo:2021wcn}, the metric satisfies Dirichlet boundary conditions, and the gravitational charges are integrable in the space of solutions. 

The linear dilaton backgrounds obtained from TsT feature nonvanishing chemical potentials inherited from AdS$_3$. The latter can be removed by large gauge transformations that guarantee the $B$-field vanishes at the origin/horizon. These large gauge transformations do not affect the conserved charges, the latter of which can be obtained from the Brown-York stress tensor. We will show that the stress tensor is finite, reproduces the charges computed in the covariant formalism, and satisfies the trace flow equation of $T\bar T$-deformed CFTs \cite{McGough:2016lol},
\eq{
T^{\mu}_{\mu} = -\frac{3\lambda}{c} \big(T_{\mu\nu} T^{\mu\nu} - T^{\mu}_{\mu} T^{\nu}_{\nu} \big),
}
where $\lambda$ is the TsT parameter and $c$ is the central charge of the undeformed CFT. 

The Euclidean on-shell action of the linear dilaton backgrounds yields the expected free energy up to the presence of the chemical potentials described earlier. When these potentials vanish, the gravitational path integral is given in the semiclassical limit by
\renewcommand{\arraystretch}{1.5}
\eqa{ \label{Zdef_intro}
\log \mathcal Z(\tau, \bar\tau,\lambda) \approx  \left\{ \begin{array}{lll} 
 i \pi  (\tau - \bar\tau) E_{\vac}(\lambda)   , & \quad & \text{ground state},\\[5pt]
- i \pi \big( \frac{1}{\tau} - \frac{1}{\bar \tau}\big) E_{\vac}\big(\frac{\lambda}{\tau\bar\tau}\big)   , & \quad\quad & \text{black hole},
\end{array} \right.
}
where $E_{\vac}(\lambda)$ is the energy of the ground state. The contributions of the ground state and black hole geometries are related by the modular $S$-transformation characteristic of $T\bar T$-deformed CFTs, namely $(\tau, \bar \tau, \lambda) \to (- 1/\tau, -1/\bar\tau, \lambda/\tau\bar\tau)$ \cite{Datta:2018thy,Aharony:2018bad}. It is then not surprising that \eqref{Zdef_intro} matches the universal partition function of (single-trace) $T\bar T$-deformed CFTs.\footnote{The gravitational path integral of three-dimensional gravity with a finite cutoff has also been shown to match the partition function of $T\bar T$-deformed CFTs \cite{Caputa:2019pam,Caputa:2020lpa,Li:2020zjb,Apolo:2023vnm}.} The matching is possible due to the form of the ground state energy which, together with modular invariance, fully determines the partition function in the semiclassical limit. Our results provide further evidence that the semiclassical limit of supergravity on the linear dilaton backgrounds possesses the essential features that characterize the $T\bar T$ deformation.\footnote{Note that in string theory, there are discrete states whose spectrum does not align with the $T\bar T$ deformation \cite{Chakraborty:2024mls}. These states are not present in the tensionless limit of the string \cite{Dei:2024sct}. Moving towards the semiclassical limit will require understanding the role of these states in the holographic correspondence.} The matching of the background energies, the trace flow equation, and the partition function are independent of the single or double-trace nature of the deformation, the latter of which is not visible within the supergravity approximation.\footnote{By the double-trace $T\bar T$ deformation we refer to the standard $T\bar T$ deformation of a CFT.}

The paper is organized as follows. In Section \ref{se:sugra}, we describe the boundary terms that must accompany the action of supergravity for the AdS$_3$ and linear dilaton backgrounds. In Section \ref{se:undef}, we evaluate the on-shell action of AdS$_3$ with arbitrary shifts of the $B$-field and show how these shifts give rise to chemical potentials in the free energy. The TsT transformation of these backgrounds is considered in Section \ref{se:tst}, where we describe the chemical potentials needed to yield the linear dilaton backgrounds associated with the $T\bar T$ deformation. In this section we compute the Brown-York stress tensor and show that it satisfies the trace flow equation of  $T\bar T$-deformed CFTs. We also compute the on-shell action, show that it matches the partition function of (single-trace) $T\bar T$-deformed CFTs, and provide an interpretation of our results.\\ %We end with our conclusions in section \ref{se:conclusions}. \\

\noindent\textbf{Note added:} During the preparation of this work, we became aware of ref.~\cite{Dei:2025ryd}, which independently computes the on-shell action of linear dilaton backgrounds using an alternative approach.

%%%%%%%%%%%%%%%%%%%%%%%%%%%%%%%%%%%%%%%%%%%%%%

%%%%%%%%%%%%%%%%%%%%%%%%%%%%%%%%%%%%%%%%%%%%%%
\section{The supergravity action} \label{se:sugra}
%%%%%%%%%%%%%%%%%%%%%%%%%%%%%%%%%%%%%%%%%%%%%%

In this section, we consider the NS sector of supergravity on spacetimes of the form $M_3 \times S^3 \times T^4$, where $M_3$ stands for an AdS$_3$ or a linear dilaton background. We describe the boundary terms necessary to obtain a finite action with a well-defined variational principle, discuss the corresponding boundary conditions, and derive a general expression for the Brown-York stress tensor.

%%%%%%%%%%%%%%%%%%%%%%%%%%%%%%%%%%%%%%%%%%%%%%
\subsection{Bulk action}
%%%%%%%%%%%%%%%%%%%%%%%%%%%%%%%%%%%%%%%%%%%%%%

In the string frame, the NS sector of supergravity dimensionally reduced on $S^3 \times T^4$ is described by the action
\eq{
  \!\! I_3 = \frac{1}{4\pi \ell_s^4} \int_{M} d^3x\sqrt{|g|} e^{-2\Phi}\bigg( R  + 4   + 4 \p_\mu \Phi \p^{\mu} \Phi  - \frac{1}{12} H_{\mu\nu\alpha} H^{\mu\nu\alpha} \bigg), \label{sugraaction3d}
}
where we have set the radius of the sphere to 1. The equations of motion are given by
\eqsp{
0&= R_{\mu\nu} - \frac{1}{2} g_{\mu\nu} R - 2 g_{\mu\nu} + 2\big (\nabla_\mu \p_\nu \Phi - g_{\mu\nu} \Box \Phi + g_{\mu\nu} \p_\alpha \Phi \p^\alpha \Phi \big) \\
& \hspace{16pt} - \frac{1}{24} \big( 6 H_{\mu \alpha \beta}H_{\nu}{}^{\alpha \beta} - g_{\mu\nu} 
H_{\alpha\beta\gamma} H^{\alpha \beta\gamma}\big), \\
0 &= \beta_\Phi  \equiv  R + 4 - 4 \p_\mu \Phi \p^\mu \Phi+ 4 \Box\Phi - \frac{1}{12} H_{\mu\nu\alpha} H^{\mu\nu\alpha} ,
 \\
0& =\p_\alpha\big(\sqrt{|g|}e^{-2\Phi}  H^{\alpha\mu\nu}\big).
}
In particular, the equation of motion of the $B$-field requires
\eq{
\sqrt{|g|} e^{-2\Phi} H^{\alpha\mu\nu} = (\text{constant}) \epsilon^{\alpha\mu\nu}.
}
This combination of terms is proportional to the electric charge, which in the ten-dimensional spacetime reads
\eq{
Q_e = \frac{1}{(2\pi \ell_s)^6} \int_{S^3 \times T^4} e^{-2\Phi_{(10)}} \star_{10} H_{(10)}.
}
This implies that the three-dimensional field strength $H$ satisfies
\eq{\label{HisQe}
\sqrt{|g|} e^{-2\Phi} H^{\alpha\mu\nu} = 2 \ell_s^2 Q_e \epsilon^{\alpha\mu\nu}, \qquad e^{-2\Phi} H_{\alpha\mu\nu} = 2 \ell_s^2 Q_e  \sqrt{|g|} \epsilon_{\alpha\mu\nu}.
}
%where $\epsilon^{uvr} = - \epsilon_{uvr} =  1$ in Lorentzian signature. In Euclidean signature we instead have $\epsilon^{\alpha\mu\nu} = \epsilon_{\alpha\mu\nu}= -i$.

On the other hand, the equation of motion of the dilaton implies that the action vanishes on shell up to a total derivative term, which follows from
\eq{\label{onshell_bulk}
I_3 = \frac{1}{4\pi \ell_s^4} \int_{M} d^3x \sqrt{|g|} e^{-2\Phi} \beta_\Phi - \frac{1}{\pi \ell_s^4} \int_{\p M} d^2x \sqrt{|\gamma|} e^{-2\Phi} n^\mu \p_\mu \Phi,
}
where $\gamma_{\mu\nu} = g_{\mu\nu} - n_\mu n_\nu$ is the induced metric at the asymptotic boundary $\p M$ and $n^\mu$ is the outward-pointing vector normal to $\partial M$. Consequently, the on-shell action of supergravity is determined entirely by boundary terms.

%%%%%%%%%%%%%%%%%%%%%%%%%%%%%%%%%%%%%%%%%%%%%%

%%%%%%%%%%%%%%%%%%%%%%%%%%%%%%%%%%%%%%%%%%%%%%
\subsection{Boundary terms}
%%%%%%%%%%%%%%%%%%%%%%%%%%%%%%%%%%%%%%%%%%%%%%

The action of supergravity must be accompanied by boundary terms that are necessary to yield finite and physically meaningful results. The most well known among these is the Gibbons-Hawking term which guarantees a well-defined variational principle for the metric. In asymptotically AdS$_3$ spacetimes, the action must also include a counterterm proportional to the area of the induced metric at the boundary, which renders the action and the Brown-York stress tensor finite \cite{Brown:1992br,Balasubramanian:1999re}. While this counterterm regularizes the action of pure three-dimensional gravity, it turns out to be insufficient in supergravity. The reason is that, in contrast to the Einstein-Hilbert action, the action of supergravity is missing the bulk term that is proportional to the volume of AdS. Therefore, the action of supergravity must be supplemented by another boundary term that depends on the $B$-field and comes with the added benefit of fixing the value of the electric charge $Q_e$. 

The total action of supergravity is proposed to be given by
\eq{
I  = I_3 + I_{GH} + 2I_{ct_1} + I_{ct_2} + I_{B}, \label{total}
}
where the Gibbons-Hawking term  $I_{GH}$, the counterterms $I_{ct_1}$ and $I_{ct_2}$, and the term $I_{B}$ are given by
\eqsp{
%\begin{gathered}
I_{GH} &= \frac{1}{2\pi \ell_s^4} \int_{\p M} d^2x \sqrt{|\gamma|} \, e^{-2\Phi} K,  \qquad I_{ct_1} = -\frac{1}{2\pi \ell_s^4} \int_{\p M} d^2x  \sqrt{|\gamma|} \, e^{-2\Phi}, \\
%= \frac{1}{8\pi \ell_s^4} \int_{\p M} d^2x e^{-2\Phi} n_\alpha H^{\alpha}{}_{\mu\nu} \epsilon^{\mu\nu}, \\
 I_{B} &= \frac{1}{4\pi \ell_s^4} \int_{\p M} e^{-2\Phi} \iota_n H \wedge \star_2 B,  \qquad   \hspace{0.2mm} I_{ct_2} = -\frac{1}{4\pi \ell_s^4} \int_{\p M} e^{-2\Phi} \iota_n H,   \label{Bterm}% = \frac{Q_e}{2\pi \ell_s^2} \int B, I_{B} = \frac{1}{8\pi \ell_s^4} \int_{\p M} d^2x \sqrt{|\gamma|} e^{-2\Phi} n_\alpha H^{\alpha \mu\nu} B_{\mu\nu},
%\end{gathered}
}
where $K = g^{\mu\nu} K_{\mu\nu} = \nabla_\mu n^\mu$ is the trace of the extrinsic curvature $K_{\mu\nu} = \frac{1}{2} \mathcal L_n \gamma_{\mu\nu}$. 

On shell, the counterterms $I_{ct_1}$ and $I_{ct_2}$ differ by a factor of $e^{-2\Phi}$ up to constants, i.e.,
\eq{
I_{ct_2, \, \text{on shell}} = \frac{Q_e}{2\pi \ell_s^2} \int_{\p M} d^2x \sqrt{|\gamma|}.
}
When the dilaton is constant, as is the case for asymptotically AdS$_3$ backgrounds, then $2I_{ct_1} + I_{ct_2} = I_{ct_1}$, and we recover the standard counterterm of pure three-dimensional gravity \cite{Balasubramanian:1999re}. However, for more general cases that include the linear dilaton backgrounds, these counterterms differ and both are necessary to obtain a finite stress tensor and a well-defined on-shell action. Altogether, the on-shell action of supergravity is a boundary term that can be written as
\eq{ \label{total_onshell}
I_{\text{on shell}} = \frac{1}{2\pi \ell_s^4} \int_{\p M}  \big( d^2x  \sqrt{|\gamma|}  \big(  \nabla_\mu (e^{-2\Phi} n^\mu) - 2 e^{-2\Phi} + \ell_s^2 Q_e \big) + \ell_s^2Q_e B \big).
}

The boundary term $I_{B}$ can be pulled into the bulk, where it reads
\eq{
I_{B,\,\text{on shell}} =  \frac{Q_e}{2\pi \ell_s^2}  \int_M  H. \label{Hterm}
}
Written in this form, $I_B$ is proportional to the one proposed by Kurlyand and Tseytlin \cite{Kurlyand:2022vzv}, who argued that the following topological term must be added to the supergravity action of AdS$_3 \times S^3 \times T^4$ spacetimes\footnote{The proposal holds more generally for spacetimes of the form $M_3 \times X^3 \times T^4$ and $M_5 \times X^5$ where $X^3$ and $X^5$ are compact.}
\eq{
\int_{\text{AdS}_3 \times S^3}  H_{\text{AdS}_3} \wedge H_{S^3}, \label{KTterm}
}
were $H_f = dB_f$ is the field strength associated with the $f = \{\text{AdS}_3, S^3\}$ factor. Being topological, \eqref{KTterm} does not affect the equations of motion. Furthermore, its on-shell value is proportional to the volume of AdS$_3$, and gives a divergent contribution to the action. After dimensional reduction on the 3-sphere, \eqref{KTterm} can be written as the boundary term \eqref{Hterm}. This boundary term was recently considered in \cite{Detournay:2024lhz}, who also noted it is necessary to obtain a finite on-shell action for asymptotically AdS$_3$ spacetimes.

Note that the bulk term \eqref{Hterm} and its boundary counterpart in \eqref{Bterm} are equivalent provided that there are no contributions of the $B$-field at the origin of coordinates. For all the solutions considered in this paper, which are independent of the non-radial coordinates, any value of the $B$-field at the origin is a constant that can be removed by a large gauge transformation. Such gauge transformations are physically meaningful, since a finite value of the $B$-field at the origin can be interpreted as turning on a chemical potential. It is in this sense that the boundary term \eqref{Bterm} is gauge invariant, as it is only sensitive to large gauge transformations which do not vanish at the asymptotic boundary. 

\subsection{The Brown-York stress tensor}

Let us now consider the variation of the action \eqref{total},
\eqa{ 
\!\!\! \delta I &= %\frac{\ell^3}{4\pi \ell_s^4} \int_\Sigma d^3x \sqrt{|G|} \text{EOM}_i \delta\psi^i +
 \frac{1}{4\pi \ell_s^4} \int_{\partial M} d^2x  \Big\{  e^{-2\Phi} \Big( \sqrt{|\gamma|}\big( K_{\mu\nu} - (K - 2 - 2 n^\alpha \partial_\alpha \Phi) \gamma_{\mu\nu} \big)  + \frac{1}{4} n_{\alpha} H^{\alpha}{}_{\rho\sigma} \epsilon^{\rho\sigma} \gamma_{\mu\nu} \Big) \delta g^{\mu\nu} \notag \\
&  \hspace{6.5pt} +  4 \sqrt{|\gamma|} \big(2 n^{\alpha} \partial_{\alpha} \Phi - K + 2 \big) e^{-2\Phi} \delta\Phi  + \frac{1}{2}\big(B_{\mu\nu} - \sqrt{|\gamma|} \epsilon_{\mu\nu}\big)\delta \big( \sqrt{|\gamma|} e^{-2\Phi} n_\alpha H^{\alpha\mu\nu} \big) \Big\} , \label{deltaI}
%&  \hspace{76pt} - \sqrt{|\gamma|}  D_\mu \big(e^{-2\Phi} n^{\alpha} \gamma^{\mu\nu} \delta g_{\nu\alpha}\big) \Big\}, 
}
where we have omitted corner terms and the bulk terms responsible for the equations of motion. In order to have a well-defined variational principle we require
\eq{\label{boundaryconditions}
\delta g_{\mu\nu}|_{\partial M} = \delta \Phi|_{\partial M} =\delta \big( \sqrt{|\gamma|} e^{-2\Phi} n_\alpha H^{\alpha\mu\nu} \big)|_{\partial M} = 0,
}
where the last term reduces to $\delta Q_e = 0$ on shell. We see that the value of the $B$-field is not fixed at the asymptotic boundary. This is a consequence of the boundary term $I_{B,\, \text{on shell}} \propto \int_{\partial M} Q_e B$, which allows us to exchange the variation of the $B$-field for the variation of the charge $Q_e$. 

The fact that the electric charge $Q_e$ must be fixed once $I_B$ is included in the action is a desirable feature. For the class of spacetimes we are interested in, $Q_e$ determines the number of fundamental strings generating the background, which in the supergravity approximation is expected to be large but fixed \cite{Maldacena:1998bw}. Relatedly, these backgrounds admit a holographic description in terms of a (deformed) CFT whose central charge is proportional to $Q_e$, which must again be large and fixed. The boundary term \eqref{Bterm} guarantees that this is always the case. Consequently, the action \eqref{total} describes the canonical ensemble with fixed temperatures and fixed charge. Omitting this term would allow us to describe a different ensemble where we fix the value of the $B$-field at the asymptotic boundary instead of $Q_e$, as described in \cite{Detournay:2024lhz}. 

The Brown-York stress tensor obtained from \eqref{deltaI} is given by
\eq{ \label{BYdef}
T_{\mu\nu} = -\frac{2}{\sqrt{|\gamma|}} \frac{\delta I}{\delta \gamma^{\mu\nu}} = -\frac{e^{-2\Phi}}{2\pi \ell_s^4} \Big( \big( K_{\mu\nu} - (K - 2 - 2 n^\alpha \partial_\alpha \Phi) \gamma_{\mu\nu} \big) + \frac{1}{4\sqrt{|\gamma|}} n_{\alpha} H^{\alpha}{}_{\rho\sigma} \epsilon^{\rho\sigma} \gamma_{\mu\nu} \Big).
}
The boundary terms added to the action guarantee that this expression is finite for both AdS$_3$ and linear dilaton backgrounds. Using the equations of motion, the Brown-York stress tensor can be written more conveniently as 
\eq{ \label{BYonshell}
T^{\text{on shell}}_{\mu\nu} = - \frac{1}{2\pi \ell_s^4} \Big( e^{-2\Phi}\big( K_{\mu\nu} - (K - 2 - 2 n^\alpha \partial_\alpha \Phi) \gamma_{\mu\nu} \big) -  \ell_s^2 Q_e \gamma_{\mu\nu} \Big).
}
For the asymptotically AdS$_3$ spacetimes considered in the next section, this version of the Brown-York stress tensor reduces to the standard one found in three-dimensional gravity. For the linear dilaton backgrounds studied in Section \ref{se:tst}, we will see that the stress tensor reproduces the gravitational charges computed in the covariant formalism.

%%%%%%%%%%%%%%%%%%%%%%%%%%%%%%%%%%%%%%%%%%%%%%

%%%%%%%%%%%%%%%%%%%%%%%%%%%%%%%%%%%%%%%%%%%%%%
\section{AdS$_3$ backgrounds} \label{se:undef}
%%%%%%%%%%%%%%%%%%%%%%%%%%%%%%%%%%%%%%%%%%%%%%

In this section, we evaluate the action of supergravity on asymptotically AdS$_3$ spacetimes. We will use the on-shell action to probe the role of large gauge transformations of the $B$-field. In particular, we will show that the boundary term $I_B$ is crucial to reproduce the action of pure three-dimensional gravity and obtain the torus partition function of holographic CFTs. In addition, we will show that this boundary term is essential to reproduce the holographic Weyl anomaly from supergravity.

%%%%%%%%%%%%%%%%%%%%%%%%%%%%%%%%%%%%%%%%%%%%%%
\subsection{The space of solutions}
%%%%%%%%%%%%%%%%%%%%%%%%%%%%%%%%%%%%%%%%%%%%%%

Let us now consider AdS$_3$ backgrounds that can be obtained from the near-horizon limit of $p$ fundamental strings and $k$ NS5-branes \cite{Maldacena:1998bw}. The family of solutions we are interested in can be conveniently written as
\eqsp{\label{undeformed}
ds^2 &= \frac{d\rho^2}{4 (\rho^2 - 4 T_u^2 T_v^2)} + \rho\, du dv + T_u^2 du^2 + T_v^2 dv^2, \\
B &= - \frac{1}{2}(\rho + 2 \bshift) du \wedge dv, \\
e^{2\Phi} &=  \frac{k}{p} ,
}
where $(u, v) = (\varphi + t, \varphi -t)$ with $\varphi \sim \varphi + 2 \pi$ and we have set the scale of AdS to $\ell = 1$. The variables $T_{u, v}$ parametrize physically distinct solutions. Here, we are interested in two classes of solutions
\eqsp{\label{adsvacuum}
\text{global AdS:} & \qquad T_{u, v} = \frac{i}{2}, \\%\qquad \quad\rho \ge \rho_0 = 1/2, \\
\text{BTZ black holes:} & \qquad T_{u,v}  \in \mathbb R_{\ge0}. %\qquad \rho \ge \rho_h = 2 T_u T_v.
}
In this gauge, the origin of coordinates $\rho_0$ and the outer/inner horizons $\rho_\pm$ of the BTZ black holes are located at 
\eq{\label{origin_horizon}
\rho_0 = -(T_u^2 + T_v^2), \qquad \rho_{\pm} = \pm 2 T_u T_v.
}

The solutions \eqref{undeformed} are characterized by additional constants that are not determined by the supergravity equations of motion. Among these, $p$ and $k$ parametrize the electric and magnetic charges of the background according to
\eq{ \label{QeQm}
Q_e = p,  \qquad Q_m = k. 
}
These constants are assumed to be fixed in the space of solutions, although in what follows it will be convenient to consider the consequences of pretending that $Q_e$ is allowed to vary. In addition, we have included an arbitrary shift of the $B$-field that is parametrized by the constant $\bshift$. This shift corresponds to a large gauge transformation which changes the chemical potential conjugate to $Q_e$, as described in more detail shortly.

The backgrounds \eqref{undeformed} are invariant under two symmetries generated by the Killing vectors $\p_u$ and $-\p_v$. The infinitesimal charges associated with these isometries can be computed using the covariant formalism \cite{Lee:1990nz,Wald:1993nt,Iyer:1994ys,Iyer:1995kg, Wald:1999wa, Barnich:2001jy,Barnich:2007bf} and are given by
\eqa{\label{deltaQ_undef}
\begin{split}
\delta E_L & \equiv \delta \mathcal Q_{\p_u \phantom{-}} = \delta\Big(\frac{c}{6} T_u^2 \Big) - k \bshift  \delta Q_e, \\
\delta E_R &\equiv \delta \mathcal Q_{-\p_v} =\delta\Big(\frac{c}{6} T_v^2 \Big) - k  \bshift  \delta Q_e,
\end{split}
}
where $c = 6k p = {3}/{2G}$ is the Brown-Henneaux central charge of the dual CFT \cite{Brown:1986nw}. The equations above show that the charges are not integrable in the space of solutions when both $\bshift$ and $Q_e$ are allowed to vary. This feature of supergravity was already noted in \cite{Apolo:2021wcn}. Note that this is not a problem for the AdS$_3$ backgrounds since, as discussed above, we must keep $Q_e$ fixed in the space of solutions such that $\delta Q_e = 0$. In this case, the charges are integrable, independent of the shift of the $B$-field, and reproduce the standard left and right-moving energies of the AdS$_3$ spacetimes in \eqref{undeformed}
\eq{\label{Q_undef}
E_L = \frac{c}{6} T_u^2, \qquad E_R  = \frac{c}{6} T_v^2.
}
In particular, the energy and angular momentum of global AdS are given by
\eq{ \label{Evac_undef}
E_{\vac} = E_L + E_R = - \frac{c}{12}, \qquad  E_L - E_R = 0.
}
%%%%%%%%%%%%%%%%%%%%%%%%%%%%%%%%%%%%%%%%%%%%%%

%%%%%%%%%%%%%%%%%%%%%%%%%%%%%%%%%%%%%%%%%%%%%%
\subsection{The chemical potential}
%%%%%%%%%%%%%%%%%%%%%%%%%%%%%%%%%%%%%%%%%%%%%%

Let us now discuss the physical consequences of the constant $\bshift$ parametrizing the $B$-field. Different choices for this constant have been considered in the literature \cite{Hemming:2001we,Troost:2002wk,Hemming:2002kd,Rangamani:2007fz,Ashok:2021ffx,Nippanikar:2021skr,Martinec:2023plo}. In particular, the choice proposed in \cite{Ashok:2021ffx,Martinec:2023plo} guarantees that the energy of winding strings matches the energy of asymptotic probes. On the other hand, it has been argued in \cite{Ashok:2021ffx,Rangamani:2007fz} that a smooth background in Euclidean signature requires the $B$-field to vanish at the origin of global AdS or the outer horizon of the BTZ black hole, case in which $\bshift$ is given by 
\eq{\label{bvalue}
\bshift_{\vac} =  T_u T_v \big|_{T_{u,v} = i/2} =  -\frac{1}{4}, \qquad \bshift_{\bh} = - T_u T_v.
}
These values of $\bshift$ guarantee that the $B$-field does not source strings in the corresponding backgrounds \cite{Martinec:2023plo}. It is interesting to note that $\bshift_{\vac}$ and $\bshift_{\bh}$ are not related by the same analytic continuation that relates the BTZ background to the global AdS one, namely\footnote{If we instead required the $B$-field to vanish at the inner horizon of the black hole, then $\bshift = T_u T_v$ would be valid for both global AdS and the BTZ black holes \cite{Hemming:2001we,Troost:2002wk}.}
\eq{
\bshift_{\vac} \ne \bshift_{\bh}\big|_{T_{u, v} = i/2}.
}

A nonvanishing $B$-field at the outer horizon of the black hole determines the chemical potential conjugate to $Q_e$. In order to see this, we first note that the left and right-moving temperatures of the black hole are given by
\eq{\label{btztemperatures}
T_L =   \frac{T_u}{\pi}, \qquad T_R  = \frac{T_v}{\pi}.
}
In terms of these variables, the entropy reads
\eq{
S = \frac{c \pi^2}{3} (T_L + T_R).
} 
It is then not difficult to show that the entropy satisfies the first law of thermodynamics, which can be written in terms of the gravitational charges as
\eq{\label{firstlaw}
\delta S = \frac{1}{T_L}  \delta E_L + \frac{1}{T_R} \delta E_R - \beta \nu_{\bh} \delta Q_e,
}
where the inverse temperature $\beta$, angular potential $\Omega$, and the chemical potential $\nu_{\bh}$ are
\eq{\label{ads_potentials}
\beta = \frac{1}{2}\Big(\frac{1}{T_L} + \frac{1}{T_R}\Big), \qquad \Omega = \frac{1}{2\beta}\Big(\frac{1}{T_R} - \frac{1}{T_L}\Big),  \qquad \nu_{\bh} = -2k(\bshift + \pi^2 T_L T_R). 
}
The first law \eqref{firstlaw} can also be derived from the gravitational charge associated with the horizon generator $\zeta_h = \p_t - \Omega \p_\varphi$ \cite{Gao:2001ut,Wald:1993nt,Jacobson:1993vj,Iyer:1994ys}. Using this approach, one finds that the chemical potential corresponds to the value of the $B$-field at the outer horizon $\rho_+$  \cite{Compere:2007vx}, in agreement with the result above
\eq{\label{nu_def}
\nu_{\bh} = k \zeta_h^\mu \xi_\varphi^\nu B_{\mu\nu}\big|_{\rho_+},
}
where for convenience we have introduced the notation
\eq{\label{xidef}
\xi_{x^\mu} = \partial_{x^\mu}.
}
Note that the chemical potential depends on the temperatures due to our parametrization of the $B$-field. A more natural choice of $B$ would make $\nu_{\bh}$ independent of the temperatures. However, the parametrization we have chosen here is useful when describing the linear dilaton backgrounds relevant in single-trace $T\bar T$ holography.
 
Additional comments are in order. First, we see that the non-integrability of the gravitational charges \eqref{deltaQ_undef} is a feature, not a bug, as it leads to a consistent first law when $Q_e$ is promoted to a variable parametrizing the space of solutions.\footnote{Note that the covariant formalism does not know about the boundary terms we have added to the action, although these can be taken into account as in \cite{Harlow:2019yfa}.} Second, if we want to have a vanishing chemical potential, the value of the $B$-field must vanish at the horizon. In this case, $\bshift$ cannot be a constant, but must depend on the parameters characterizing the solution according to \eqref{bvalue}, such that the $B$-field becomes
\eq{
B = - \frac{1}{2}(\rho -2 T_u T_v) du\wedge dv.
}
When $\delta Q_e = 0$, the fact that $\bshift$ depends on the variables parametrizing the solution is not an issue. As described earlier, in this case the gravitational charges \eqref{deltaQ_undef} are always integrable,  independent of the choice of $\bshift$, and given by \eqref{Q_undef}.  The reason we stress this is that, as we will see in Section \ref{se:tst}, this will no longer be generally true for the linear dilaton backgrounds obtained from TsT transformations unless $\bshift$ takes particular values. 
 
In Section \ref{se:onshell_undef}, we will use the on-shell action to verify that the chemical potential conjugate to $Q_e$ is indeed given by \eqref{nu_def}. This will allow us to extend the definition of the chemical potential to the global AdS case, which we will show can be obtained from the value of the $B$-field at the origin.
%%%%%%%%%%%%%%%%%%%%%%%%%%%%%%%%%%%%%%%%%%%%%%

%%%%%%%%%%%%%%%%%%%%%%%%%%%%%%%%%%%%%%%%%%%%%%
\subsection{The holographic Weyl anomaly}
%%%%%%%%%%%%%%%%%%%%%%%%%%%%%%%%%%%%%%%%%%%%%%

Before we turn to the evaluation of the on-shell action, let us make some comments about the holographic Weyl anomaly. In pure three-dimensional gravity, the Weyl anomaly can be obtained from the bulk part of the on-shell action \cite{Henningson:1998gx}. In contrast, in supergravity, the Weyl anomaly originates from the $B$-field via the boundary term \eqref{Bterm}. This can be seen by noting that in the Fefferman-Graham gauge, locally AdS$_3$ backgrounds satisfy
\eq{
ds^2 = \frac{dr^2}{r^2} + r^2  g^{(0)}_{\mu\nu} + g^{(2)}_{\mu\nu} + \frac{1}{r^2} g^{(4)}_{\mu\nu}, \qquad B = -\Big(\int dr \sqrt{|g|}\,\Big) du \wedge dv, \qquad \Phi = \text{const}, \label{AAdS}
}
where $g^{(2)}_{\mu\nu}$ and $g^{(4)}_{\mu\nu}$ are given by
\eq{
g^{(0)\mu\nu} g^{(2)}_{\mu\nu} = -\frac{1}{2} R^{(0)}, \qquad g^{(4)}_{\mu\nu} = \frac{1}{4} g^{(2)}_{\mu\alpha} g^{(0)\alpha\beta} g^{(2)}_{\beta\nu}.
}
As a result, the expansion of the $B$-field features a logarithmic divergence that is proportional to the Ricci scalar $R^{(0)}$ of the boundary metric $g^{(0)}_{\mu\nu}$, that is
\eq{
B  = - \frac{1}{2} \sqrt {|g^{(0)}|} \Big( r^2 +\frac{1}{2} R^{(0)}  \log r + \mathcal O(r^0) \Big) du \wedge dv.
}
The on-shell action of the asymptotically AdS$_3$ backgrounds \eqref{AAdS} is then given by
\eq{
I_{\text{on shell}} = -\int_{\p M} d^2x \sqrt{\big|g^{(0)}\big|} \Big( -\frac{c}{24\pi} R^{(0)} \Big) \log \epsilon + \mathcal O(r^0), 
}
where $\epsilon \to 0$ is a UV cutoff. From this expression we can extract the Weyl anomaly
\eq{
\mathcal A = -\frac{c}{24\pi} R^{(0)}.
}
The result above is not surprising, since the bulk version of the boundary term $I_B$, when evaluated on shell, is nothing but the volume of the spacetime, that is
\eq{
 I_{B,\,\text{on shell}} =  \frac{Q_e}{2\pi \ell_s^2} \int_M H\big|_{\text{on shell}}  =  \frac{1}{16\pi G} \int_M d^3x \sqrt{|g|}(R + 2 )\big|_{\text{on shell}}.
}

%%%%%%%%%%%%%%%%%%%%%%%%%%%%%%%%%%%%%%%%%%%%%%

%%%%%%%%%%%%%%%%%%%%%%%%%%%%%%%%%%%%%%%%%%%%%%
\subsection{The on-shell action} \label{se:onshell_undef}
%%%%%%%%%%%%%%%%%%%%%%%%%%%%%%%%%%%%%%%%%%%%%%

Let us now evaluate the on-shell action of supergravity on the Euclidean continuation of the AdS$_3$ backgrounds \eqref{undeformed}. First, we note that for these backgrounds, the counterterms in \eqref{Bterm} are proportional to each other such that
\eq{
2I_{ct_1} + I_{ct_2} = I_{ct_1} = -\frac{1}{8\pi G} \int_{\p M} d^2x \sqrt{|\gamma|}.
}
Consequently, the Brown-York stress tensor \eqref{BYdef} reduces to the standard stress tensor of three-dimensional gravity, namely \cite{Balasubramanian:1999re}
\eq{
T_{\mu\nu} = -\frac{1}{8\pi G} \big( K_{\mu\nu} - (K -1) \gamma_{\mu\nu} \big).
}
The stress tensor is traceless and yields the same gravitational charges obtained from the covariant formalism \eqref{Q_undef}. Note also that, as described earlier, the term $I_B$ in \eqref{Bterm} is not invariant under large gauge transformations of the $B$-field. This means, in particular, that $I_B$ depends on the choice of $\bshift$ parametrizing shifts of the $B$-field. When the dilaton is a constant independent of the phase space variables $T_{u,v}$, constant shifts of the $B$-field do not affect the gravitational charges or the first law of thermodynamics. The story is different, however, when it comes to the on-shell action. Indeed, we will see that in order to reproduce the action of pure three-dimensional gravity, it is necessary to choose the values of $\bshift$ given in \eqref{bvalue}.

The Euclidean continuation of \eqref{undeformed} can be obtained by letting $t \to i t_E$, whereupon the $u$ and $v$ coordinates become complex conjugates of one another. In addition, we compactify the time coordinate such that the Euclidean spaces have the topology of filled-in tori satisfying
\eq{\label{taudef}
u \sim u + 2 \pi \sim u + 2\pi \tau, \qquad (\tau, \bar \tau) = \bigg(\frac{i}{2\pi T_L}, - \frac{i}{2\pi T_R}\bigg). %= \frac{\tilde\Omega + i \beta}{2\pi}.
}
%where $(\tilde\Omega, \beta) = (- i \beta \Omega,\beta)$. 
The Euclidean action is given in terms of the Lorentzian one by $I_E = - I$. Thus, the gravitational path integral is given in the semiclassical approximation by
\renewcommand{\arraystretch}{1.5}
\eqa{ \label{Zundef}
\!\! \log \mathcal Z(\tau, \bar\tau) \approx - I_E = \left\{ \begin{array}{lll} 
i \pi  (\tau - \bar\tau) E_{\vac} -  i  \pi (\tau - \bar\tau)   \nu_{\vac} Q_e, & & \text{thermal AdS},\\[5pt]
 - i \pi \big( \frac{1}{\tau} - \frac{1}{\bar \tau}\big) E_{\vac}  +  i  \pi \big(\frac{1}{\tau} -  \frac{1}{\bar\tau} \big) |\tau|^2  \nu_{\bh} Q_e , & \phantom{ii} & \text{BTZ black hole},
\end{array} \right.
}
where $E_{\vac}$ is the ground state energy \eqref{Evac_undef} and the chemical potentials $\nu_{\vac}$, $\nu_{\bh}$ satisfy
\eqsp{\label{nu_AdS}
\nu_{\vac} &= k \xi_t^\mu \xi_\varphi^\nu B_{\mu\nu}\big|_{\rho_0}   = -2 k\bigg( \bshift + \frac{1}{4}\bigg) , \\ % = -2 \bshift - \frac{1}{2} , \\ %=  - \frac{1}{2}(1 + 4 \bshift ), \\
\nu_{\bh} &=  k \zeta_h^\mu \xi_\varphi^\nu B_{\mu\nu}\big|_{\rho_+} =  -2k(\bshift + \pi^2 T_L T_R). % -2k \big( \bshift + T_u T_v\big) = -2 \bshift - 2 T_u T_v. %=  - 2(T_u T_v +  \bshift).
}
In particular,  it is not difficult to verify that the free energy is given by
\eq{
\beta F = I_E = \frac{1}{T_L} E_L + \frac{1}{T_R} E_R - S - \beta \nu Q_e.
}
The on-shell action of the thermal AdS and BTZ backgrounds is finite and, modulo the chemical potentials, reproduces the results expected from pure three-dimensional gravity and holography \cite{Hartman:2014oaa}. An exact match is obtained when the chemical potentials vanish, namely, when the values of the $B$-field at the origin of thermal AdS and the horizon of the BTZ black hole vanish.

%%%%%%%%%%%%%%%%%%%%%%%%%%%%%%%%%%%%%%%%%%%%%%

%%%%%%%%%%%%%%%%%%%%%%%%%%%%%%%%%%%%%%%%%%%%%%
\section{Linear dilaton backgrounds} \label{se:tst}
%%%%%%%%%%%%%%%%%%%%%%%%%%%%%%%%%%%%%%%%%%%%%%

In this section we consider the linear dilaton backgrounds obtained from TsT transformations of the AdS$_3$ solutions considered in the previous section. We will show that in order to recover the entropy of $T\bar T$-deformed CFTs, it is necessary to set $\bshift = 0$ in the undeformed $B$-field. We will compute the on-shell actions of these backgrounds and show that they reproduce the torus partition function of (single-trace) $T\bar T$-deformed CFTs up to chemical potentials conjugate to $Q_e$. These chemical potentials do not affect the entropy or the gravitational charges, and can be turned off by a large gauge transformation that ensures the $B$-field vanishes at the origin/horizon. The gravitational charges can be computed using the Brown-York stress tensor, which reproduces the expressions obtained from the covariant formalism, and is shown to satisfy the trace flow equation of $T\bar T$-deformed CFTs.

%%%%%%%%%%%%%%%%%%%%%%%%%%%%%%%%%%%%%%%%%%%%%%
\subsection{The space of solutions via TsT}
%%%%%%%%%%%%%%%%%%%%%%%%%%%%%%%%%%%%%%%%%%%%%%

Let us now perform a TsT transformation of the AdS$_3$ backgrounds \eqref{undeformed} consisting of a T-duality along $u$, a shift $v \to v  - \lambda u$, and another T-duality along $u$. In order to keep track of the effects of the $B$-field, we perform a shift $B \to B - \Bshift du \wedge dv$ after the TsT transformation. The resulting linear dilaton backgrounds are given by\footnote{It would be interesting to understand if more general backgrounds satisfying the same boundary conditions can be constructed using the asymptotic $T$-dualities introduced in \cite{Detournay:2024lhz}.}
  \eqsp{\label{deformed}
   ds^2 &= \frac{d\rho^2}{ 4 \big (\rho^2 - 4 T_u^2 T_v^2 \big )} + \frac{\rho \, du dv+T_u^2 du^2 + T_v^2 dv^2}{h(\rho)} ,  \\
   B &=  - \frac{1}{2}  \bigg( \frac{\rho + 2 \l T_u^2 T_v^2 + 2 (1 + \lambda \rho) \bshift + 2 \lambda \bshift^2}{2h(\rho)} + 2 \Bshift \bigg) \,du \wedge dv, \\ 
  e^{2\Phi} &=   \frac{k h(\rho)}{\eta p}  ,
  }
where $h(\rho)$ and $\eta$ read
\eq{ \label{heta_def}
h(\rho) = \frac{1}{1 + (2 \bshift + \rho) \lambda + ( \bshift^2 + T_u^2 T_v^2 + \bshift \rho)\lambda^2 }, \qquad \eta = \frac{1}{1 + 2 \bshift  \lambda +  (\bshift^2 - T_u^2 T_v^2)\lambda^2}.
}
The background \eqref{deformed} takes the same form obtained in \cite{Apolo:2019zai} except that $h(\rho)$ and $\eta$ now depend on the shift $\bshift$ of the undeformed $B$-field according to \eqref{heta_def}. The dependence of the dilaton on $h(\rho)$ is determined by Buscher's rule \cite{Buscher:1987sk,Buscher:1987qj} while the constant $\eta$, which is not determined by this rule, is related to the field strength $H$ by
\eq{
H = - \frac{h(\rho)^2}{2 \eta} du \wedge dv \wedge d\rho.
}
This choice of the dilaton guarantees that the electric and magnetic charges of the deformed background are fixed and given by their undeformed values \eqref{QeQm}. 

The shift of the $B$-field before the TsT transformation affects the boundary conditions of the deformed background. In the region $\rho \to \infty$, the metric and $B$ field satisfy
\eq{\label{bc_def}
ds^2 \big|_{\rho \to \infty} \sim \frac{dr^2}{4r^2} + \frac{du dv}{\lambda(1 +\bshift \lambda )} , \qquad B \big|_{\rho \to \infty} \sim -\frac{1}{2}  \bigg(\frac{1 + 2\bshift  \lambda  }{\lambda(1 + \bshift\lambda )}  + 2\Bshift \bigg) du \wedge dv.
}
The dependence of the asymptotic value of the $B$-field on $\bshift$ and $\Bshift$ is not a problem since we are fixing the conjugate charge $Q_e$ instead. However, the shift $\bshift$ of the undeformed $B$-field does affect the boundary conditions of the metric. In particular, for the values of $\bshift$ for which the undeformed chemical potential vanishes \eqref{bvalue}, the asymptotic value of the metric would depend on the variables parametrizing the space of solutions. In other words, the metric would no longer satisfy the Dirichlet boundary conditions required for a well-defined variational principle. In this case, either the space of solutions must be restricted or the action must be supplemented by additional boundary terms to accommodate these boundary conditions.

\paragraph{The ground state.} 
In analogy with the undeformed case, we are interested in two classes of solutions, the ground state and rotating black holes. The ground state can be obtained by the analytic continuation 
\eq{\label{ancont}
T_{u, v} = i \sigma, \qquad \sigma \in \mathbb R.
}
Expanding the metric and $B$-field near the origin via $\rho \to 2 \sigma^2 (1 + 2 \epsilon^2)$ we find\footnote{The origin of coordinates and the location of the outer and inner horizons of the black hole are unchanged by the TsT transformation and given by \eqref{origin_horizon}.}
\eqsp{
ds^2 &\sim d\epsilon^2 + 4\sigma^2h(2\sigma^2) \epsilon^2  d\varphi^2 - 4\sigma^2 h(2\sigma^2) dt^2 + \dots, \\
B &\sim  - \big( h(2\sigma^2)^{1/2}\,(\bshift + \sigma^2) + \Bshift\big) du \wedge dv + \dots.
}
A smooth background solution with a vanishing $B$-field at the origin is obtained by requiring
\eq{\label{vacuum_condition}
4 \sigma^2 h(2\sigma^2) = \frac{4\sigma^2}{(1 + \lambda (\bshift + \sigma^2))^2} = 1, \qquad  \Bshift  + \frac{\bshift + \sigma^2}{1 + \lambda (\bshift + \sigma^2)} = 0.
}
The solution to the first equation depends on the functional dependence of $\bshift$ on the $T_{u,v}$ parameters. Here we will consider two cases. The first case is $\bshift = 0$, for which we recover the energy of the ground state of (single-trace) $T\bar T$-deformed CFTs. We postpone the discussion of this case to the end of this section. Another interesting case is found when $\bshift = \bshift_{\vac} = -1/4$, namely, when the chemical potential of the undeformed background vanishes. In this case, the solution to \eqref{vacuum_condition} with a well defined $\lambda \to 0$ limit is given by
\eq{ \label{groundstateNotTTbar} 
\bshift =\bshift_{\vac} \quad \implies \quad T_{u, v} = i \sigma = \frac{i}{2}, \quad \Bshift = 0.
}
This is the same analytic continuation as that of the undeformed background \eqref{adsvacuum}, which differs from (and is much simpler than) the one obtained in the $T\bar T$ case where $\bshift = 0$~\eqref{vacuum_TTbar}.  In this case, the TsT transformation maps global AdS to the ground state of a theory whose ground-state energy differs from that of a $T\bar T$-deformed CFT. This contrasts with the $\bshift =0$ case, where the preimage of the deformed ground state does not coincide with the undeformed one, a mismatch that persists for any state other than the massless black hole. Instead, in order to map an undeformed solution with parameters $T_{u,v}$ to the corresponding $T\bar T$-deformed one, it is necessary to reparametrize the $T_{u,v}$ variables, as described in \cite{Apolo:2019zai}.  We interpret this feature as a consequence of the nonvanishing chemical potential present before the TsT transformation.

\paragraph{The black holes.} 
The other class of solutions we are interested in are black holes described by \eqref{deformed} with $T_{u,v} \in \mathbb R_{\ge 0}$. The near-horizon limit of the black hole solutions can be obtained by letting $\rho  \to 2 T_u T_v(1 + 2 \epsilon^2)$ and is described by
\eqsp{\label{near-horizon}
ds^2 &\sim  d\epsilon^2 - \epsilon^2 h(2T_u T_v) (T_u du - T_v dv)^2 + h(2T_u T_v) (T_u du + T_v dv)^2 + \dots, \\
B &\sim - \big( h(2T_u T_v)^{1/2}\,(\bshift + T_u T_v) + \Bshift \big) du \wedge dv + \dots.
}
The periodicities of the coordinates $(u, v) \sim (u + i /T_L, v - i/ T_R)$ required for a smooth Euclidean continuation yield the deformed left and right-moving temperatures
\eqsp{ \label{TLTR_def}
T_L &= \frac{1}{\pi} \sqrt{h(2T_u T_v)} \, T_u= \frac{1}{\pi} \frac{T_u}{1 + \lambda (\bshift  + T_u T_v)}, \\
T_R &=  \frac{1}{\pi} \sqrt{h(2T_u T_v)} \, T_v  = \frac{1}{\pi} \frac{T_v}{1 + \lambda (\bshift + T_u T_v)}.
}
Furthermore, a vanishing value of the $B$-field at the horizon requires
\eq{
\Bshift =  -  \sqrt{h(2T_u T_v)} \, \big(\bshift + T_u T_v\big) = - \frac{\bshift + T_u T_v}{1 + \lambda(\bshift + T_u T_v)}.
}
Interestingly, when the undeformed chemical potential $\nu_{\bh}$ vanishes, namely when $\bshift = \bshift_{\bh} = - T_u T_v$, the temperatures and chemical potential of the black hole are unchanged by the TsT transformation, i.e.,
\eq{
\bshift = \bshift_{\bh} \quad \implies \quad T_L = \frac{T_u}{\pi}, \quad T_R = \frac{T_v}{\pi}, \quad \Bshift = 0.
} 
This feature of TsT transformations was noticed previously in \cite{Detournay:2012dz}. As with the case of the ground state with $\bshift = \bshift_{\vac}$, this means that the TsT transformation maps a BTZ black hole at temperatures $T_{L,R}$ to a deformed black hole at the same temperatures. However, as we can see here, this map can be affected by large gauge transformations of the $B$-field prior to the TsT transformation.

The entropy of the black hole can be obtained from the area of the horizon in Planck units. From the near-horizon metric \eqref{near-horizon} we find that the entropy is given by 
\eq{\label{TsT_entropy}
S = \frac{\pi c \eta}{3 \sqrt{h(2 T_u T_v)}}(T_u + T_v) = \frac{\pi c}{3} \frac{T_u + T_v}{1 + \lambda ( \bshift - T_u T_v)} .
}
There is only one choice of $\bshift$ for which the entropy takes the form characteristic of  $T\bar T$-deformed CFTs,
\eq{
S_{T\bar T} = \frac{\pi^2 c}{3}\bigg(\frac{T_L + T_R}{\sqrt{1 - 4\pi^2\lambda T_L T_R}} \bigg) .
}
Using \eqref{TLTR_def}, we find that the entropy \eqref{TsT_entropy} matches the entropy of $T\bar T$-deformed CFTs only when $\bshift = 0$, namely\footnote{For completeness, we note that when the undeformed chemical potential vanishes ($\bshift = \bshift_{\bh}$), the entropy is given in terms of the temperatures $T_{L,R}$ by
\eq{
S\big|_{\bshift = \bshift_{\bh}} = \frac{\pi^2 c}{3} \bigg( \frac{T_L + T_R}{1 - 2 \pi^2\lambda T_L T_R}\bigg), \notag
}
which is manifestly not the formula expected for $T\bar T$-deformed CFTs.}
\eq{
 S_{T \bar T}= \frac{\pi c }{3}\bigg(\frac{T_u + T_v}{\sqrt{(1 + \lambda(\bshift - T_u T_v))^2 + 4 \bshift \lambda^2 T_u T_v}} \bigg)  \quad \underset{\bshift \to 0}{\longrightarrow} \quad S\big|_{\bshift = 0} .
}
Therefore, if we are interested in the space of solutions associated with the $T\bar T$ deformation, we must set $\bshift = 0$ as originally done in \cite{Apolo:2019zai}. This case will be considered in more detail towards the end of this section.

\paragraph{Conserved charges.} 
Let us now compute the conserved charges of the TsT-transformed backgrounds \eqref{deformed} using the covariant formalism. In this approach, it is necessary to know how $\bshift$ and $\Bshift$ change in the space of solutions of the theory. Assuming that $\bshift$ and $\Bshift$ are arbitrary parameters, we find that the infinitesimal charges associated with the Killing vectors $\partial_u$ and $-\partial_v$ are given by
\eqa{\label{deltaQ_def}
\begin{split}
\delta E_L & \equiv \delta \mathcal Q_{\p_u \phantom{-}} = \delta\Big(\frac{c \eta}{6} \big(T_u^2( 1 + \lambda T_v^2) - \lambda \bshift^2\big) \Big) - \frac{c}{6} \bshift \,\delta \eta - \eta k  (\Bshift + \eta \bshift) \,\delta Q_e, \\
\delta E_R &\equiv \delta \mathcal Q_{-\p_v} = \delta\Big(\frac{c \eta}{6} \big(T_v^2( 1 + \lambda T_u^2) - \lambda \bshift^2\big) \Big) - \frac{c}{6} \bshift \,\delta \eta - \eta k (\Bshift + \eta\bshift) \,\delta Q_e.
\end{split}
}
These expressions feature two sets of terms that are generically not integrable and because of which the charges would not be well defined.\footnote{The non-integrable terms have a common origin, namely, the shift of the dilaton by a variable that changes in the space of solutions.} The last term in these expressions is not physical since, as discussed earlier, we have $\delta Q_e = 0$; nevertheless keeping this term will be useful in determining the chemical potential of the black hole.  Note that in this case ($\delta Q_e = 0$), the charges are independent of the large gauge transformations of the $B$-field parametrized by $\Bshift$. 

The term proportional to $\bshift \,\delta \eta$ in \eqref{deltaQ_def} is not manifestly integrable unless $\delta \bshift = 0$ or $\bshift$ depends on $T_{u,v}$ in a particular way. This can be seen from
\eq{\label{int_test}
\bshift \,\delta \eta = - \frac{\lambda}{\big(1 + 2\lambda \bshift + \lambda^2(\bshift^2 - T_u^2 T_v^2)\big)^2} \big( \delta \bshift^2 + \lambda \bshift \, \delta( \bshift^2 - T_u^2 T_v^2)\big).
}
This expression is integrable whenever $\delta \bshift = 0$ and in particular when $\bshift = 0$. When $\delta \bshift \ne 0$, a necessary condition for the integrability of \eqref{int_test} is that $\bshift$ is a function of $T_u T_v$. This is precisely the case for the values of $\bshift$ that make the chemical potentials of global AdS and the BTZ black holes vanish \eqref{bvalue}. In these cases we have $\bshift = \pm T_u T_v$ and \eqref{int_test} becomes
\eq{ \label{int_beta}
\bshift \,\delta \eta \big|_{\bshift = \pm T_u T_v} = - \frac{2\lambda T_u T_v}{(1 \pm 2\lambda T_u T_v )^2} \delta (T_u T_v) = \pm \frac{1}{2\lambda} \bigg(\frac{2\lambda T_u T_v}{1 \pm 2\lambda T_u T_v} \mp \log\big(1 \pm 2 \lambda T_u T_v\big) \bigg).
}
This expression is interesting (and troubling) because of the logarithm, which suggests that $\bshift \to \pm T_u T_v$ is not compatible with the shift of the dilaton that keeps $Q_e$ fixed. This is not a problem for the TsT backgrounds associated with the $T\bar T$ deformation for which $\bshift = 0$.

\paragraph{The chemical potential.} 
We now consider the first law of thermodynamics for the black holes. From the entropy \eqref{TsT_entropy} and the infinitesimal charges \eqref{deltaQ_def} we find that
\eq{\label{firstlaw_def}
\delta S = \frac{1}{T_L}  \delta E_L + \frac{1}{T_R} \delta E_R - \beta \nu \delta  Q_e,
}
where $\beta$ is the inverse temperature and the deformed chemical potential $\nu$ reads
\eq{
\nu(\lambda) = -2k\bigg(\Bshift + \frac{\bshift + T_u T_v}{1 + \lambda( \bshift + T_u T_v) }\bigg) .
}
The presence of a chemical potential for the TsT-transformed backgrounds was already noted in \cite{Apolo:2019zai} for the case $\bshift = \Bshift = 0$. Note, however, that \eqref{firstlaw_def} holds more generally for any value of the parameters characterizing the solutions in \eqref{deformed}. As expected, we find that despite the complicated and non-integrable form of the conserved charges \eqref{deltaQ_def}, the latter are compatible with the first law of thermodynamics.

The first law \eqref{firstlaw_def} can also be derived using the covariant formulation of charges, as described earlier for the BTZ black holes. In this approach, the chemical potential is given by the value of the $B$-field at the horizon, which matches the expression reported above. Similarly, we expect the chemical potential of the ground state to be given by the value of the $B$-field at the origin, as verified earlier for global AdS. The chemical potentials for the TsT-transformed backgrounds are proposed to be given by
\eqsp{\label{nuTsT}
\nu_{\vac} (\lambda) &= k\xi_t^\mu \xi_\varphi^\nu B_{\mu\nu}\big|_{\rho_0}  = - 2k \bigg( \Bshift + \frac{\bshift - T_u T_v}{1 + \lambda( \bshift - T_u T_v) }\bigg) \bigg|_{T_{u,v} = i \sigma},  \\
\nu_{\bh} (\lambda)&=  k \zeta_h^\mu \xi_\varphi^\nu B_{\mu\nu}\big|_{\rho_+}  = -2k \bigg( \Bshift + \frac{\bshift + T_u T_v}{1 + \lambda( \bshift + T_u T_v) } \bigg) .
}
We observe that in the limit $\lambda \to 0$, these chemical potentials reduce to the undeformed values given in \eqref{nu_AdS} with $\Bshift + \bshift$ playing the role of $\bshift$. In particular, this shows that the chemical potentials of the linear dilaton backgrounds are the result of nonvanishing chemical potentials in the AdS$_3$ solutions prior to the TsT transformation.

\paragraph{The space of solutions for $T\bar T$.} 
We have seen that TsT transformations are capable of reproducing the ground state and black hole solutions associated with the $T\bar T$ deformation only when $\bshift =0$, in agreement with the results of \cite{Apolo:2019zai}. For this reason, we will set $\bshift = 0$ for the rest of the paper. We stress that in this case, the charges of the TsT-transformed backgrounds are integrable and independent of $\Gamma$. The value of $\Bshift$ can be fixed by requiring the $B$-field to vanish at the origin of the ground state and the horizon of the black hole geometries. In the next section, we will show that this is necessary to reproduce the partition function of $T\bar T$-deformed CFTs.

Starting with the undeformed AdS$_3$ backgrounds \eqref{undeformed}, the backgrounds associated with the $T\bar T$ deformation can be generated via the following prescription
\eq{
\text{shift of $B$-field by $-\bshift$} \quad + \quad \text{TsT by $\lambda$} \quad + \quad \text{shift of $B$-field by $\Bshift$}.
}
The ground state and black hole solutions are characterized by\\[2pt]
{\it\underline{Ground state}:}
\eqsp{\label{vacuum_TTbar}
\begin{gathered}
T_{u, v} = i \sigma = i\bigg[\frac{1}{\lambda} \big(1 - \sqrt{1 - \lambda}\big) \bigg]^2, \qquad \nu_{\vac}(\lambda) = -2 k\bigg( \Bshift + \frac{ 1 - \sqrt{1- \lambda} }{2\lambda} \bigg) , \\
E_{\vac}(\lambda) = E_L + E_R = -\frac{c}{6 \lambda} (1 - \sqrt{1 - \lambda}\,), \qquad E_L - E_R = 0.
\end{gathered}
}
{\it\underline{Black holes}:}
\eqa{
T_L &=  \frac{1}{\pi} \frac{T_u}{1 + \lambda   T_u T_v}, \qquad  T_R = \frac{1}{\pi} \frac{T_v}{1 + \lambda  T_u T_v},  \qquad \nu_{\bh}(\lambda) =  -2 k \bigg( \Bshift  +  \frac{1 - \sqrt{1 - 4\pi^2 \lambda T_L T_R}}{2\lambda} \bigg) , \notag \\
& \hspace{64pt} E_L(\lambda) = \frac{c}{6}\frac{(1 + \lambda T_v^2)T_u^2}{1 - \lambda^2 T_u^2 T_v^2} , \qquad E_R(\lambda) =  \frac{c}{6}\frac{(1 + \lambda T_u^2)T_v^2}{1 - \lambda^2 T_u^2 T_v^2}. \label{blackholes_TTbar}
}
The ground state energy in \eqref{vacuum_TTbar} matches the energy of the ground state in single and double-trace $T\bar T$-deformed CFTs, after an appropriate identification of parameters, as described in more detail in the next section. Similarly, the black hole energies \eqref{blackholes_TTbar} match the thermal expectation values of the energies in single and double-trace $T\bar T$ deformed CFTs, which can be understood as a consequence of modular invariance, see Section \ref{se:tst_discussion}. 
%%%%%%%%%%%%%%%%%%%%%%%%%%%%%%%%%%%%%%%%%%%%%%

%%%%%%%%%%%%%%%%%%%%%%%%%%%%%%%%%%%%%%%%%%%%%%
\subsection{The trace flow equation} \label{se:tst_traceflow}
%%%%%%%%%%%%%%%%%%%%%%%%%%%%%%%%%%%%%%%%%%%%%%

The Brown-York stress tensor defined in \eqref{BYdef} is finite for all of the linear dilaton backgrounds given in \eqref{deformed}. When $\gamma = 0$, the stress tensor reads
\renewcommand{\arraystretch}{2.25}
\eq{\label{BY_tst}
T_{\mu\nu} = \frac{c}{12\pi}\left( \begin{array}{ccc}
    \dfrac{T_u^2}{1- \lambda^2 T_u^2 T_v^2} & \phantom{i} &  \dfrac{\lambda T_u^2 T_v^2}{1- \lambda^2 T_u^2 T_v^2}  \\
    \dfrac{\lambda T_u^2 T_v^2}{1- \lambda^2 T_u^2 T_v^2}  &&  \dfrac{T_v^2}{1- \lambda^2 T_u^2 T_v^2} 
\end{array} \right).
}
The stress tensor gives us an alternative way to compute the gravitational charges, which match the results obtained from the covariant formalism in \eqref{vacuum_TTbar} and \eqref{blackholes_TTbar}, namely
\eq{
E_L =  \int_0^{2\pi} d\varphi  \, T_{\mu\nu} \, \xi_t^\mu \xi_u^\nu, \qquad E_R =  \int_0^{2\pi} d\varphi  \,  T_{\mu\nu} \, \xi_t^\mu \xi_{-v}^\nu.
}
Note that the trace of the stress tensor does not vanish, as expected for a dual field theory without conformal invariance. In fact, the stress tensor satisfies the trace flow equation characteristic of $T\bar T$-deformed CFTs
\eq{ \label{traceflow}
T^{\mu}_{\mu} = -\frac{3\lambda}{c} \mathcal O_{T\bar T}, \qquad \mathcal O_{T\bar T} =  T_{\mu\nu} T^{\mu\nu} - T^{\mu}_{\mu} T^{\nu}_{\nu} ,
}
where the trace and the contractions of the stress tensor are taken using the Minkowski metric given by the induced metric at the boundary \eqref{bc_def} after a rescaling by $\lambda$.

The trace flow equation \eqref{traceflow} can be interpreted from the point of view of the single or double-trace $T\bar T$ deformations, depending on the identification of parameters. Let us first consider the double-trace case, for which the trace flow equation is more readily matched to its field theory counterpart. For the double-trace deformation of a CFT with central charge $c$ and deformation parameter $\mu$, the holographic dictionary is given by
\eq{ \label{dictionary_doubletrace}
\text{\it\underline{Double trace\vphantom{g}}:} \qquad \mu = \frac{3\lambda}{c}.
}
This identification of variables guarantees that \eqref{traceflow} reproduces the trace flow equation of $T\bar T$-deformed CFTs \cite{McGough:2016lol}. Moreover, it ensures that the ground state and black hole energies, as well as the gravitational path integral, match the corresponding quantities in $T\bar T$-deformed CFTs, as discussed in more detail in the next section.

The Brown-York stress tensor cannot distinguish between the single or double-trace nature of the $T\bar T$ deformation. The string theory origin of the linear dilaton backgrounds favors a single-trace interpretation at the tensionless limit, but the situation is less clear in the semiclassical regime, as discussed in Section \ref{se:tst_discussion}. Nevertheless, for the single-trace deformation of a symmetric product orbifold CFT with $p$ copies, seed central charge $c_0$, and deformation parameter $\mu_0$, the holographic dictionary deduced from string theory reads~\cite{Apolo:2019zai}
\eq{ \label{dictionary_singletrace}
\text{\it\underline{Single trace}:} \qquad c = p c_0 , \qquad c_0 = 6k, \qquad \mu_0 = \frac{3\lambda}{c_0} .
}

The matching of the trace flow equation in the single-trace case is a more delicate task. In order to see this, let us denote the stress tensor and the $T\bar T$ operator in the $i$th copy of the symmetric orbifold by $T_{\mu\nu}^{(i)}$ and $\mathcal O^{(i)}_{T\bar T}$, respectively. The total stress tensor and the single-trace $T\bar T$ operator of the theory are then given by
\eq{
T_{\mu\nu} = \sum_{i=1}^p T^{(i)}_{\mu\nu}, \qquad \tilde{\mathcal O}_{T\bar T} = \sum_{i=1}^p \mathcal O^{(i)}_{T\bar T}.
}
In the single-trace deformation, each copy of the symmetric orbifold is deformed by the $T\bar T$ operator in that copy. It follows that the trace of the stress tensor in the $i$th copy satisfies its own trace flow equation, namely
\eq{
T^{(i)}{}^{\mu}_{\mu} = - \mu_0  \mathcal O^{(i)}_{T\bar T}.
}
As a result, the trace of the total stress tensor of the symmetric orbifold does not satisfy \eqref{traceflow} since it is the single trace operator $\tilde{\mathcal O}_{T\bar T}$, instead of the double-trace one $\mathcal O_{T\bar T}$, that appears on the right-hand side 
\eq{\label{singletrace_traceflow}
T^{\mu}_{\mu} =  - \mu_0  \tilde{\mathcal O}_{T\bar T}.
}
Nevertheless, this trace flow equation can still be matched to \eqref{traceflow}. 

The key insight that allows us to match \eqref{traceflow} and \eqref{singletrace_traceflow} is that the linear dilaton backgrounds can be interpreted as states in single-trace $T\bar T$-deformed CFTs whose energies are distributed uniformly among all of the copies of the symmetric orbifold. This is clearly the case for the ground state, but it is also true for thermal states, since a uniform distribution of the energies maximizes the entropy \cite{Giveon:2017nie,Apolo:2019zai}.\footnote{This implies that each untwisted state contributes a $1/p$ fraction of the total energy, while each state with twist $w$ contributes a $w/p$ fraction of the total energy.} We interpret this result as the statement that, in the semiclassical limit, the stress tensor from each copy of the symmetric orbifold contributes a $1/p$ fraction of the total stress tensor, that is
\eq{
T^{(i)}_{\mu\nu} = \frac{1}{p} T_{\mu\nu}. 
}
Consequently, the trace flow equation for this type of states --- which are the typical states of symmetric orbifolds at large temperature and large central charge --- satisfy
\eq{
T^{\mu}_{\mu} =  - \mu_0  \tilde{\mathcal O}_{T\bar T} = - \frac{\mu_0}{p} \mathcal O_{T\bar T}.
}
Using the dictionary \eqref{dictionary_singletrace}, we see that this expression reproduces the trace flow equation of the linear dilaton backgrounds \eqref{traceflow}.
%%%%%%%%%%%%%%%%%%%%%%%%%%%%%%%%%%%%%%%%%%%%%%

%%%%%%%%%%%%%%%%%%%%%%%%%%%%%%%%%%%%%%%%%%%%%%
\subsection{The on-shell action} \label{se:tst_onshell}
%%%%%%%%%%%%%%%%%%%%%%%%%%%%%%%%%%%%%%%%%%%%%%

We now compute the on-shell action of the linear dilaton backgrounds associated with the single-trace $T\bar T$ deformation. One of our goals will be to verify the values of the chemical potentials of the ground state and black hole solutions. We will also show that when these chemical potentials vanish, the on-shell actions of the ground state and the black hole reproduce the torus partition function of (single-trace) $T\bar T$-deformed CFTs.

We begin by noting that the action \eqref{total}, reproduced below for convenience,
\eq{
I = I_3 + I_{GH} + 2I_{ct_1} + I_{ct_2} + I_B,
} 
is not invariant under TsT transformations. The $I_3 + I_{GH}$ terms and the $I_{ct_1}$ counterterm are independently invariant under TsT transformations, as a consequence of Buscher's rule. This can be seen by writing these terms as follows
\eq{
I_3 + I_{GH} =  \frac{1}{2\pi \ell_s^4} \int_{\p M}  \sqrt{\frac{\gamma}{g}} \, \partial_\mu \big( \sqrt{|g|} e^{-2\Phi} n^\mu \big), \qquad I_{ct_1} = -\frac{1}{2\pi \ell_s^4} \int_{\p M}  \sqrt{|\gamma|} e^{-2\Phi}. \label{TsTinvariants}
}
It is important to keep in mind that the TsT-transformed backgrounds \eqref{deformed} have an additional shift of the dilaton parametrized by $\eta$, meaning that the expressions in \eqref{TsTinvariants} are given by $\eta$ times their undeformed values. On the other hand, the counterterm $I_{ct_2}$ and the term $I_B$ are not TsT invariant due to their dependence on the $B$-field. This is a desirable feature in fact, since a TsT-invariant on-shell action would be independent of the deformation parameter $\lambda$, and would not be able to reproduce the torus partition function of $T\bar T$-deformed CFTs.\footnote{Note that the aforementioned shift of the dilaton, which would affect a TsT-invariant action, depends on $\lambda$ but would be insufficient to reproduce the partition function of $T\bar T$-deformed CFTs.} In fact, it is difficult to construct other TsT invariant boundary terms that are local, diff invariant, and do not spoil the boundary conditions. 

Let us now evaluate the on-shell action on the Euclidean continuation of the linear dilaton backgrounds with $\bshift = 0$. As in the undeformed case, we perform the analytic continuation $t \to i t_E$ and identify $u \sim u + 2 \pi \sim u + 2\pi \tau$ where $\tau$ is given in \eqref{taudef}. In the gauge used in this paper, the bulk $I_3$ and boundary terms $I_{GH}$ and $2I_{ct}$ are independently divergent, but their combination vanishes
\eq{
I_3 + I_{GH} + 2I_{ct}  = 0.
}
As mentioned above, this combination of terms is TsT invariant, meaning that this is the same result obtained from the AdS$_3$ backgrounds (up a factor of $\eta$ which does not change the result). On the other hand, the counterterm $I_{ct_2}$ and the term $I_{B}$ are finite and read
\eq{\label{Ict2IB}
I_{ct_2} = \frac{c \beta}{6\lambda} , \qquad I_{B} = - \frac{c \beta}{6\lambda} - \frac{c\beta \Bshift}{3} .
}

The term $I_B$ is sensitive to large gauge transformations of the $B$-field, which explains its dependence on $\Bshift$. The boundary terms \eqref{Ict2IB} are not TsT invariant and acquire a nontrivial dependence on the deformation parameter $\lambda$. Interestingly, both of these terms have an ill-defined $\lambda \to 0$ limit, although their combination is finite. The reason is that the $\rho \to \infty$ and $\lambda \to 0$ limits do not commute, as can be seen from the asymptotic values of the metric and the $B$-field in \eqref{bc_def}. One may wonder how is it possible to recover the undeformed values of $I_{ct_2}$ and $I_B$ from the expressions above. This can be achieved if the $\lambda \to 0$ limit is taken in two steps that are reminiscent of the cutoff approach to the $T\bar T$ deformation \cite{McGough:2016lol}. Indeed, if we first let $\lambda \to \epsilon$, where $\epsilon$ is the UV cutoff of the theory, and then take the $\epsilon \to 0$ limit, we recover the undeformed values of $I_{ct_2}$ and $I_B$ (with $\Bshift \to \bshift$).

Altogether, the Euclidean on-shell action of the linear dilaton backgrounds with $\bshift = 0$ is given by\footnote{For completeness, we note that the on-shell action of the linear dilaton backgrounds with arbitrary values of $\bshift$ is also finite and given by
\eqst{
I_E = \frac{c\beta}{3} \bigg( \Bshift + \frac{\bshift}{1 + \lambda \bshift}\bigg).
}
}
\eq{\label{IE1}
I_E = \frac{c \beta}{3} \Bshift.
}
While finite, this action seems to be far from the contributions expected from the partition function of $T\bar T$-deformed CFTs. The reason is that the TsT transformation produces backgrounds with particular chemical potentials. In order to see this, we first note that
\eqsp{\label{onshell1}
i \pi( \tau - \bar\tau) E_{\vac}(\lambda) &= \frac{c\beta}{6\lambda} \big(1 - \sqrt{1 - \lambda}\,\big),  \\
-i \pi \bigg(\frac{1}{\tau} - \frac{1}{\bar\tau}\bigg) E_{\vac}\big(\tfrac{\lambda}{\tau\bar\tau}\big) &= \frac{c\beta}{6\lambda} \big(1 - \sqrt{1 - 4\pi^2 \lambda T_L T_R}\big) ,  
}
where $E_{\vac}(\lambda)$ is the energy of the ground state \eqref{vacuum_TTbar}. On the other hand, the chemical potentials in \eqref{vacuum_TTbar} and \eqref{blackholes_TTbar} yield
\eqsp{ \label{onshell2} 
- i \pi(\tau - \bar\tau)  \nu_{\vac}(\lambda) Q_e &=  -\frac{c\beta}{6\lambda}  \big(  1 - \sqrt{1- \lambda} + 2 \lambda\Bshift \big),\\
i \pi \bigg(\frac{1}{\tau} - \frac{1}{\bar\tau}\bigg)   |\tau|^2\nu_{\bh}(\lambda) Q_e   & = -  \frac{c\beta}{6\lambda} \big(  1 - \sqrt{1 - 4\pi^2 \lambda T_L T_R}  + 2 \lambda\Bshift   \big) .
}
Thus, the gravitational path integral can be written in the semiclassical approximation as
\renewcommand{\arraystretch}{1.5}
\eqa{ \label{Zdef}
\! \log \mathcal Z(\tau, \bar\tau,\lambda) \approx  \left\{ \begin{array}{lll} 
 i \pi  (\tau - \bar\tau) E_{\vac}(\lambda) -  i  \pi (\tau - \bar\tau)  \nu_{\vac}(\lambda) Q_e , &  & \text{ground state},\\[5pt]
- i \pi \big( \frac{1}{\tau} - \frac{1}{\bar \tau}\big) E_{\vac}\big(\frac{\lambda}{\tau\bar\tau}\big)  +  i  \pi \big(\frac{1}{\tau} -  \frac{1}{\bar\tau} \big) |\tau|^2 \nu_{\bh}(\lambda) Q_e   , &  \phantom{ii} & \text{black hole}.
\end{array} \right.
}
The contributions of the ground state and the black hole are related by the modular $S$-transformation characteristic of $T\bar T$-deformed CFTs \cite{Datta:2018thy,Aharony:2018bad},
\eq{\label{modularTTbar}
(\tau, \bar\tau,\lambda) \to  \biggl(- \frac{1}{\tau},  - \frac{1}{\bar\tau},  \frac{\lambda}{\tau \bar\tau} \bigg),
}
provided that we make an appropriate choice of $\Bshift$ for the black hole solutions. In particular, we find that the free energy takes the expected form, namely
\eq{
\beta F = I_E = \frac{1}{T_L} E_L + \frac{1}{T_R} E_R - S - \beta \nu Q_e.
}

Let us now compare the gravitational result \eqref{Zdef} to the partition function of $T\bar T$-deformed CFTs. As mentioned earlier, the TsT transformation of the AdS$_3$ backgrounds produces solutions with nontrivial gravitational charges, temperatures, and chemical potentials. However, TsT transformations are not sufficient to describe the space of solutions associated with the $T\bar T$ deformation. As pointed out in \cite{Apolo:2019zai}, an additional shift of the dilaton is required to keep the electric charge fixed. Moreover, turning off the chemical potentials that accompany the TsT transformation requires an additional shift of the $B$-field. In analogy with the undeformed AdS$_3$ backgrounds, the shift of the $B$-field takes different functional forms for the ground state and black hole solutions. As a result, these shifts are not related by the same analytic continuation that relates one background to the other. From \eqref{vacuum_TTbar} and \eqref{blackholes_TTbar} we see that the shifts of the $B$-field are given by
\eqsp{
\Bshift_{\vac} &= \frac{T_u T_v}{1 - \lambda T_u T_v} \bigg|_{T_{u,v} = i \sigma} =  - \frac{1 - \sqrt{1 - \lambda}}{2 \lambda}, \\
\Bshift_{\bh} &= \frac{ -T_u T_v}{1 + \lambda T_u T_v}  =  - \frac{1 - \sqrt{1 - 4\pi^2 \lambda T_L T_R}}{2 \lambda}.
}
For these choices, the chemical potentials vanish and the contributions to the gravitational path integral \eqref{Zdef} reproduce the torus partition function of $T\bar T$-deformed CFTs in the semiclassical regime \cite{Apolo:2023aho},
\eq{ \label{Z_TTbar}
\log Z(\tau, \bar\tau, \mu) \approx \max\bigg\{ \pi i (\tau - \bar\tau) \mathcal E_{\vac}(\mu), \, - i \pi \Big(\frac{1}{\tau} - \frac{1}{\bar\tau}\Big) \mathcal E_{\vac}\Big(\frac{\mu}{\tau\bar\tau}\Big) \bigg\}, \qquad |\tau|^2 \ne 1,
}
where $\mathcal E_{\vac}(\mu)$ denotes the ground state energy of the $T\bar T$-deformed CFT. 

The on-shell action of supergravity is not able to distinguish between the single and double-trace $T\bar T$ deformations, meaning that the matching to the partition function \eqref{Z_TTbar} is valid for both single and double-trace $T\bar T$-deformed CFTs. In the single-trace case, the ground state is given by the product of the ground states of each copy in the symmetric orbifold. Thus, the holographic dictionary \eqref{dictionary_singletrace} shows that the total energy satisfies
\eqsp{
&\text{\it\underline{Single trace}:} \qquad \mathcal E_{\vac}(\mu_0) =  - \frac{p}{2\mu_0} \bigg(1 - \sqrt {1 - \frac{ c_0 \mu_0}{3} }  \, \bigg) = E_{\vac}(\lambda).
}
For the double-trace deformation \eqref{dictionary_doubletrace}, we similarly find
\eq{
\text{\it\underline{Double trace\vphantom{g}}:} \qquad  \mathcal E_{\vac}(\mu) =  - \frac{1}{2\mu} \bigg(1 - \sqrt {1 - \frac{ c \mu}{3} }  \, \bigg) = E_{\vac}(\lambda).
}
In both cases, the total central charge appearing in the deformed partition function is $c = p c_0$ and the deformation parameters satisfy
\eq{
c \mu = c_0 \mu_0 = 3 \lambda.
}
%In other words, the supergravity computation is not sensitive to quantities that would allow us to distinguish between the single and double-trace $T\bar T$ deformations. For this, one needs string theory. 
%Note that the thermal expectation values of the energies derived from \eqref{Z_TTbar} match the deformed energies of $T\bar T$-deformed CFTs and reflect the generalized modular invariance of the partition function. For this reason, the energies computed in supergravity cannot distinguish between the single and double-trace $T\bar T$ deformations. 
The situation here is similar to the one encountered in the undeformed case, where the on-shell action of the AdS$_3$ backgrounds reproduces the torus partition function of holographic CFTs in the semiclassical limit, and is not able to distinguish whether these CFTs arise from a symmetric product orbifold. In fact, the insensitivity of gravity to the nature of the dual CFT extends to thermal two-point functions, as described recently in \cite{Belin:2025nqd}.
%%%%%%%%%%%%%%%%%%%%%%%%%%%%%%%%%%%%%%%%%%%%%%

%%%%%%%%%%%%%%%%%%%%%%%%%%%%%%%%%%%%%%%%%%%%%%
\subsection{Discussion} \label{se:tst_discussion}
%%%%%%%%%%%%%%%%%%%%%%%%%%%%%%%%%%%%%%%%%%%%%%

The string theory origin of the linear dilaton backgrounds would favor the single-trace interpretation of our results --- or a deformation of it, as described shortly --- but it is possible that a holographic description exists that is independent of string theory, and for which the single-trace picture is not necessary. Here we wish to emphasize that, rather than the matching to the partition function of single or double-trace $T\bar T$-deformed CFTs, the agreement essentially follows from the form of the ground state energy and the generalized modular invariance characteristic of $T\bar T$-deformed CFTs. In particular, modular invariance fixes the asymptotic growth of states, justifying the expression for the entropy, as well as the expectation values of the left and right-moving energies, expressions that match either the single or double-trace formulae, as described in \cite{Apolo:2023aho}.  

The single-trace version of $T\bar T$ holography is expected to be valid in the tensionless limit of string theory where $k = 1$. Concrete evidence for this expectation, beyond the one provided for arbitrary values of $k$ in \cite{Giveon:2017nie, Hashimoto:2019wct, Apolo:2019zai, Hashimoto:2019hqo, Georgescu:2022iyx,Apolo:2023aho,Chakraborty:2023wel,Cui:2023jrb,Du:2024bqk}, has been given in \cite{Dei:2024sct}, which matches the one-loop torus partition function at the tensionless limit. In order to move away from the stringy regime where a supergravity description is not valid, the single-trace theory must be deformed by additional operators. In analogy with \cite{Gaberdiel:2015uca,Apolo:2022fya,Fiset:2022erp}, these operators must break the symmetric orbifold structure and lift the Hagedorn density of light states of these theories, the latter of which is not compatible with a semiclassical gravitational description. 

The results of this paper support the idea that the deformed theory dual to the semiclassical limit of supergravity on the linear dilaton backgrounds must still be a $T\bar T$-deformed CFT in a very concrete sense: its partition function must be invariant under the modular transformation~\eqref{modularTTbar}.\footnote{This statement is valid in the semiclassical approximation where the contribution of matter fields and string excitations is suppressed. In the string theory setup of \cite{Giveon:2017nie}, discrete states are expected to modify the modular properties of the partition function, and it will be important to understand the role of these states in the holographic correspondence.} In addition, the stress tensor of the deformed theory must satisfy the trace flow equation characteristic of the $T\bar T$ deformation \eqref{traceflow}. Finally, as shown in \cite{Georgescu:2022iyx}, we also expect the deformed theory to preserve the nonlocal symmetries of the $T\bar T$ deformation \cite{Guica:2019nzm,Guica:2021pzy,Guica:2020uhm,Guica:2022gts}. The generalized modular invariance and nonlocal symmetries of $T\bar T$-deformed CFTs may be helpful in characterizing the deformation of the single-trace theory away from the symmetric orbifold point where the single-trace version of $T\bar T$ holography is more firmly established. We hope to return to this question in the future.

\vfill

%%%%%%%%%%%%%%%%%%%%%%%%%%%%%%%%%%%%%%%%%%%%%%
\section*{Acknowledgements}
%%%%%%%%%%%%%%%%%%%%%%%%%%%%%%%%%%%%%%%%%%%%%%
 
It is a pleasure to thank Alejandra Castro, Ben Freivogel, Dominik Neuenfeld, Andrew Rolph, Wei Song, and Jan Troost for helpful discussions. This work was supported by the Beijing Natural Science Foundation International Scientist Project No.~IS24015. 

%%%%%%%%%%%%%%%%%%%%%%%%%%%%%%%%%%%%%%%%%%%%%%

\vfill

%%%%%%%%%%%%%%%%%%%%%%%%%%%%%%%%%%%%%%%%%%%%%%
%\subsection*{Acknowledgments}
%%%%%%%%%%%%%%%%%%%%%%%%%%%%%%%%%%%%%%%%%%%%%%

%%%%%%%%%%%%%%%%%%%%%%%%%%%%%%%%%%%%%%%%%%%%%%

\bibliographystyle{JHEP}
\bibliography{refs}

\end{document}
%%%%%%%%%%%%%%%%%%%%%%%%%%%%%%%%%%%%%%%%%%%%%%
% The end
%%%%%%%%%%%%%%%%%%%%%%%%%%%%%%%%%%%%%%%%%%%%%%